# Temperature Effect on Phonon Dispersion Stability of Zirconium by Machine Learning-driven Atomistic Simulations

Xin Qian and Ronggui Yang[*]

Department of Mechanical Engineering

University of Colorado, Boulder, CO 80309, USA

Email: Ronggui.Yang@Colorado.Edu

ORCID: 0000-0002-3198-2014 (Xin Qian)

0000-0002-3602-6945 (Ronggui Yang)

**Abstract**

It is well known that conventional harmonic lattice dynamics cannot be applied to dynamically unstable crystals at 0 K, such as high temperature body centered cubic (BCC) phase of crystalline Zr. Predicting phonon spectra at finite temperature requires the calculation of force constants to the third, fourth and even higher orders, however, it remains challenging to determine to which order the Taylor expansion of the potential energy surface for different materials should be cut off. Molecular dynamics, on the other hand, intrinsically includes arbitrary orders of phonon anharmonicity, however, its accuracy is severely limited by the empirical potential field used. Recently, machine learning algorithms emerge as a promising tool to build accurate potentials for molecular dynamics simulation. In this work, we approach the problem of predicting phonon dispersion at finite temperature by performing molecular dynamics simulations with machine learning-driven potential fields. We developed Gaussian approximation potential models for both the hexagonal closed-packed (HCP) phase and the body centered cubic (BCC) phase of Zirconium crystals. The developed potential field is first validated with static properties including energy-volume relationship, elastic constants and phonon dispersions at 0 K. Molecular dynamics simulations are then performed to stochastically sample the potential energy surface and to calculate the phonon dispersion at elevated temperatures. The phonon renormalization in BCC-Zr is successfully captured by the molecular dynamics simulation at 1188 K. The instability of BCC structure is found to originate from the double-well shape of the potential energy surface where the local maxima is located in an unstable equilibrium position. The stabilization of the BCC phase at high temperature is due to the dynamical average of the low-symmetry minima of the double well due to atomic vibrations.

# 1. INTRODUCTION

Understanding temperature-dependent thermal properties of materials is important for a lot of high temperature applications, such as thermal barrier coatings, nuclear applications and high temperature thermoelectrics. Prediction of macroscopic thermal properties depend on the microscopic description of vibrational dynamics of the atoms in the solids, which is primarily characterized by phonon dispersions. Although recent progress in first-principles calculation has enabled prediction of thermal properties routinely for many materials, it has been one of the long-standing challenges in material physics to model the vibrational spectra for materials that are dynamically unstable. Conventionally, lattice dynamics calculations are performed at the static limit (0 K) using the finite displacement method[1] or density functional perturbation theory[2], but these methods failed to explain why the dynamically unstable structures can emerge at high temperatures. For example, SnSe in the *CmCm* phase is one of thermoelectric materials with best figure of merit ZT at the high temperature (~1000 K). However, the *CmCm* structure displays soft phonon modes with imaginary frequencies in the phonon dispersion at the static limit. For these soft phonons, the harmonic force constants are negative, which means that the inter-atomic forces no longer pull the atoms back to the equilibrium position but push them away once the atoms are displaced from the equilibrium position. Clearly, the existence of soft phonons is a sign of lattice instability, but the static lattice dynamics failed to explain why the *CmCm* phase of SnSe is stable at high temperature. Another example is the body centered cubic (BCC) structure for group IV metals like Ti, Zr and Hf. They all have soft phonons at the static limit but become stable phases at high temperature.[3,4] In 1955, Hooton realized that atoms vibrate in an effective potential due to their nonstationary neighbors, and the potential energy surface (PES) is stochastically sampled around the most probable position which is not necessarily a local minima.[5] They then renormalized the soft phonon modes by an effective harmonic potential that is temperature-dependent. Along this line, the problem of dynamical instability is addressed by a self-consistent approach under the harmonic approximation,[6] which starts with the phonon dispersion at static limit as an initial guess and iteratively solve the eigenmodes of the dynamical equation. However, several recent studies suggest that care must be taken for strongly anharmonic crystals where the PES should be expanded to the third and even the fourth order.[7-9] Therefore, the accuracy of the force constants could be significantly affected by the artificial truncation of the Taylor expansion of the PES.[8] On the other hand, Classical molecular dynamics can naturally incorporate the phonon anharmonicity of arbitrary order without truncating the Taylor expansion of the PES, but it suffers from the inaccuracy of the empirical potential field as limited by the fitting with the empirical functional forms.[10-12]

To overcome the challenges of both the first-principles lattice dynamics and the molecular dynamics simulations using empirical potential, machine learning (ML) based regression algorithms provide an elegant solution to reconstruct the *ab-initio* PES. Instead of decomposing the PES to simple empirical functional forms, the ML algorithm is totally data-driven, which fits the PES by "learning" the correlation between the atomic configurations and the resulting energy from the *ab-initio* data.[13] Since the ML algorithm does not assume any form of functions when fitting the *ab-initio* PES, it does suffer from the error caused by artificially truncating the Taylor expansions of the PES. In principle, the ML algorithm includes all orders of anharmonic terms in the PES. Such data-driven feature of ML algorithms also resulted in a significantly improved accuracy of the ML-based potential compared with the empirical potentials, because it bypasses the difficulty of decomposing the high dimensional PES to simple functional forms when fitting for empirical potentials. Due to these advantages, machine learning algorithms including artificial neural networks[14], Gaussian process regression, [15] and others[16] have been successfully used to model the thermal and mechanical properties in simple crystals such as Si, [17-19] GaN,[18] and graphene,[20] as well as complex atomistic structures and processes, such as the amorphous carbon,[21] lithium ion transport in electrode materials, [22, 23] and phase-change material GeTe[24].

Since machine learning algorithms addressed both the problem of truncating expansions of PES in first-principles calculations and the inaccuracy problem of the empirical potentials, it could be a promising tool to capture the lattice dynamics above 0 K by fitting the PES at elevated temperatures. This paper is therefore focused on modeling the phonon renormalization using ML-driven potential in Zirconium (Zr) crystal, one of the most classic example of dynamical instability. Zr and its alloys are indeed widely used as cladding materials in nuclear reactors.[25] At room temperature, Zr takes the hexagonal closed packed (HCP) phase and transitions into a body centered cubic (BCC) phase at higher temperature, which is dynamically unstable at 0 K.[6] Since phase stability is usually required to prevent structural failures in nuclear applications, understanding the temperature dependent vibrational dynamics of elemental Zr is critical. Recently, Zong et al. successfully reproduced the phase diagram of Zr using a potential developed by kernel ridge regression algorithm,[26] indicating that ML could be a promising tool to model lattice dynamics of dynamically unstable crystals. However, their potential has limited accuracy for predicting phonon dispersion of both HCP and BCC Zr, with discrepancy of optical phonon frequency as large as 2 THz at the Brillouin zone center.[26] This is probably because their machine learning potential was developed to reproduce phase diagram based on a multi-phase-learning strategy. The training database therefore contains multi-phase structures with regions of phase space beyond thermal vibrations, which is unnecessary for modeling phonons. As a result, the accuracy of phonon dispersions could be compromised.[11] It remains unexplored whether such ML potential can be applied to study the temperature-induced renormalization of the soft phonon modes in dynamically unstable structures.

In this paper, we focused on modeling the temperature effect on phonon dispersions using ML potential. Gaussian approximation potential (GAP) model[18, 27] based on the Gaussian Process Regression algorithm[15] is used to fit the PES of both HCP-Zr and BCC-Zr. For each phase of Zr, we developed a GAP model which accurately reproduced the energies and interatomic forces, the equation of state and the elastic constants derived from first-principles calculations. We observed that the instability of the BCC Zr at the static limit originates from the double-well shape of the PES, and the BCC structure corresponds to the local maxima of the PES. The high temperature BCC structure is stabilized by a stochastic average due to atomic vibrations over the two low symmetry minima separated by a low potential barrier. The phonon renormalization of the BCC-Zr can therefore be captured by performing molecular dynamics (MD) simulations which stochastically samples the PES. Using spectral energy density analysis, we have successfully observed that the soft transverse acoustic (TA) phonons of BCC-Zr is renormalized to ~ 1 THz at 1188 K.

## 2. METHODOLOGY

Here we briefly review the formalism to use the GAP method for fitting PES and the symmetry invariant descriptors for characterizing the atomic configurations in Section 2.A. We then discuss the details for generating the database from the first-principles calculations including total energies, inter-atomic forces and virial stresses for training the machine learning based GAP model in Section 2.B. The training databases are downloadable in supplementary materials,[28] and the training process is performed using the QUIP package.[29]

## A. Fitting Potential Energy Surface using GAP Method

To construct the machine learning-driven potential using GAP, the total energy of the simulation cell is decomposed into the contributions from each individual atom:

$$E = \sum_i \varepsilon(\boldsymbol{q}_i) \quad (1)$$

where $\varepsilon(\boldsymbol{q}_i)$ is the contribution of energy from atom $i$, and $\boldsymbol{q}_i$ is the descriptor vector that characterizes the local chemical environment of atom $i$, i.e. the configurations of atoms in the neighborhood of atom $i$. The local energy contribution $\varepsilon(\boldsymbol{q})$ is given by a linear combination of the kernel functions:

$$\varepsilon(\boldsymbol{q}_i) = \sum_j \alpha_j K(\boldsymbol{q}_i, \boldsymbol{q}_j) = \sum_j K_{ij}\alpha_j \quad (2)$$

where the summation over $j$ includes all the atomic configurations in the first-principles database. The kernel function $K_{ij} = K(\boldsymbol{q}_i, \boldsymbol{q}_j)$ is a nonlinear function that quantifies the degree of similarity between the chemical environments described by $\boldsymbol{q}_i$ and $\boldsymbol{q}_j$. The vector $\boldsymbol{\alpha} = (\alpha_1, \alpha_2, \ldots, \alpha_j, \ldots)$ are the unknown coefficients to be determined using the first-principles data. Here we discuss first the determination of the unknown coefficient vector $\boldsymbol{\alpha}$, which is also called as "training process", and then briefly discuss the specification of the kernel function $K$ and descriptors $\boldsymbol{q}_i$. Detailed derivations can be found in refs [30, 31].

The database for building the GAP potential is collected into the vector $\boldsymbol{y}$, which contains the results from the first-principles calculations including total energies, inter-atomic forces and virial stresses. Another vector $\boldsymbol{\varepsilon}$ is introduced to denote the set of local atomic energies with components $\varepsilon_j = \varepsilon(\boldsymbol{q}_j)$. Then a linear operator $\boldsymbol{L}$ can be introduced to correlate $\boldsymbol{y}$ and $\boldsymbol{\varepsilon}$ through $\boldsymbol{y} = \boldsymbol{L}^\mathrm{T}\boldsymbol{\varepsilon}$. The operator $\boldsymbol{L}$ is then constructed as follows. If the data entry $y_i$ in vector $\boldsymbol{y}$ is the total energy of a certain atomic configuration, then $(L^T)_{ij}$ is 1 if the local energy $\varepsilon(q_j)$ of atom $j$ should be included into the summation to find total energy as shown in Eq. (1), otherwise $(L^T)_{ij}$ is 0. If the data $y_i$ is a component of interatomic forces or stresses, then $(L^T)_{ij}$ are differential operators $\frac{\partial}{\partial x_j}$ with respect to atomic coordinate $x_j$. Using the linear operator $L$, the covariance matrix $K_{DD}$ can be constructed to quantify the similarity correlation between any pair of data points in the vector $y$ as:

$$K_{DD} = L^T K_{NN} L \tag{3}$$

where the subscript $D$ and $N$ denotes the length of $y$ and $\varepsilon$, respectively. $K_{NN}$ is the covariance matrix for the joint covariance matrix for energies with elements $(K_{NN})_{ij} = K(q_i, q_j)$ corresponding to the atomic configurations in $\varepsilon$. However, computing the full covariance matrix $K_{NN}$ is expensive since $N$ can easily approach $10^5$ when the forces and virial stresses are included in the database. Therefore, a sparsification method[30] is used to reduce the computational cost. Instead of computing the full matrix $K_{NN}$, a representative set containing $M$ atoms ($M \ll N$) are chosen from the full set of $N$ atoms randomly, so that the computational cost is reduced by dealing with a much smaller covariance matrix $K_{MN}$ between the representative set and the full set and the covariance $K_{MM}$ of the representative set. Then the unknown coefficients $\alpha = (\alpha_1, \alpha_2, \ldots, \alpha_M)^T$ is calculated as a linear combination of the input data $y$, which is derived from Bayesian probability formula:

$$\alpha = (K_{MM} + K_{MN} L \Lambda^{-1} L^T K_{MN}^T)^{-1} K_{MN} L \Lambda^{-1} y \tag{4}$$

where $\Lambda$ is a diagonal matrix with diagonal elements the squared uncertainties ($\sigma_v^2$) of the input data due to convergence parameters in *ab-initio* calculations, see Table I.

We now discuss the formalism for the kernel functions $K(q, q')$ and the descriptor vector $q$. The descriptor vector $q$ is used to characterize the structural features of atomic configurations in the neighborhood of a certain atom (later referred as local chemical environments), which is usually referred as the chemical environment. The descriptor of a dimer molecule is simply the bond length between the two atoms. However, in condensed matter systems like crystals, one needs to deal with the many body feature of atomic interactions, which makes the choice of descriptor much more difficult. One of the most intuitive choice of descriptor for solids is the list of atomic positions $\{r_i\}_{i=1}^N$. However, $\{r_i\}_{i=1}^N$ is not a good descriptor, because it fails to uniquely characterize certain atomic configurations. For example, one can simply generate a complete different list by changing the order of atoms in the list, or imposing arbitrary rotations/translations to the coordinates, while the new list and the old list corresponds to the same atomic structure. A good descriptor should therefore be invariant to permutation, translation and rotation operations.[27] Recently, Bartok *et al.* derived the so called SOAP descriptor[32] that can be used to uniquely characterize and differentiate chemical environments, which is chosen as the descriptor in this work. Since the nonlocal metallic bonds in Zr crystals are intrinsically many-body interactions, the many-body SOAP descriptor becomes the natural choice. In SOAP, the chemical environment of an atom $i$ is represented by the density of neighboring atoms, which is smoothed by a Gaussian function:

$$\rho_i(r) = \sum_j e^{-\frac{|r-r_{ij}|^2}{2\sigma_a^2}} f_{cut}(|r_{ij}|) \qquad (5)$$

where $r_{ij} = r_i - r_j$ is the vector connecting atom $i$ and its neighboring atom $j$, $\sigma_a$ is corresponding to "size" of atom. The function $f_{cut}$ is a smooth cut-off function:

$$f_{cut}(r) = \begin{cases} 1, & r < r_{cut} - d \\ \frac{1}{2}\left[1 + \cos\left(\pi \frac{r - r_{cut} + d}{d}\right)\right], & r_{cut} - d < r \leq r_{cut} \\ 0, & r > r_{cut} \end{cases} \quad (6)$$

where $r_{cut}$ is the cutoff radius, and $d$ is the cutoff transition width where the $f_{cut}$ smoothly decays from 1 to 0. Obviously, $\rho_i$ only depends on the relative coordinate $\mathbf{r}_{ij}$ thus invariant to translations, and the summation over $j$ ensured permutation invariance of $\rho_i$. To ensure the rotational invariance, the atomic density distribution $\rho_i$ is further expanded to a set of orthonormal radial basis functions $g_n(r)$ and spherical harmonics $Y_{lm}$:

$$\rho_i(\mathbf{r}) = \sum_{n<n_{max}} \sum_{l<l_{max}} \sum_{m=-l}^{l} c_{nlm}^i g_n(|\mathbf{r}|) Y_{lm}\left(\frac{\mathbf{r}}{|\mathbf{r}|}\right) \quad (7)$$

The components in descriptor vector $\mathbf{q}_i$ are then calculated as the power spectrum of the expansion coefficients $c_{nlm}^i$:

$$(\mathbf{q}_i)_{nn'l} = \sum_m \left(c_{nlm}^i\right)^* c_{n'lm}^i \quad (8)$$

After specifying the descriptors, the kernel functions are constructed by inner products of descriptor vectors:

$$K_{ij} = \sigma_w^2 \left|\frac{\mathbf{q}_i \cdot \mathbf{q}_j}{|\mathbf{q}_i| \cdot |\mathbf{q}_j|}\right|^\zeta \quad (9)$$

where the exponent $\zeta$ is a positive integer to improve the sensitivity to different local atomic environments, and $\sigma_w^2$ is an overall scaling parameter. From Eq.(5) to Eq. (9), the hyper parameters $(\sigma_v, \sigma_a, \sigma_w, r_{cut}, d, \zeta, n_{max}, l_{max})$ are summarized in Table I for constructing the descriptors and the kernel functions. Here we choose typical values of $\sigma_a, \sigma_w, d, r_{cut}, \zeta$ in the literature.[31, 33] The expansion cutoff $n_{max}, l_{max}$ are chosen so that a converged phonon dispersion can be obtained with the tolerance in frequency of 0.01 THz.

In summary, the procedure of fitting PES works as follows. The data from first-principles calculations are collected into the vector $y$ first and the coefficient vector $\alpha$ is then calculated using Eq. (4). The kernel functions used to generate covariance matrices $K_{MN}, K_{MM}$ are specified as Eq (5-9). After obtaining the coefficient vector $\alpha$, total energies of an arbitrary atomic configuration $q$ can be calculated using Eq (1-2), which completes the Gaussian process regression process. In the following part, we are going to discuss the details for generating the training database, *i.e.*, the vector $y$ using the first-principles calculations.

Table I. Hyper parameters for GAP with SOAP kernels.

| | |
|---|---|
| $r_{cut}$ | 5.0 Å |
| $d$ | 1.0 Å |
| $\sigma_v$ for energy | 0.001 eV/atom |
| $\sigma_v$ for forces | 0.05 eV/Å |
| $\sigma_v$ for virial stress | 0.05 eV/atom |
| $\sigma_w$ | 1.0 eV |
| $\sigma_a$ | 0.5 Å |
| $\zeta$ | 4 |
| $n_{max}$ | 12 |
| $l_{max}$ | 12 |

### B. Generation of Training Database

Since the purpose of this work is to model the temperature effect on phonon dispersion of Zr, the database should be constructed with specific emphasis on the phase space region around equilibrium that is approachable by thermal vibrations. The developed potential is expected to accurately fit the curvature of *ab-initio* PES at equilibrium. In addition to the curvature at the static limit, the thermal vibrations would sample a wider region of the PES in the phase space, which is essentially the physical origin for phonon dispersion renormalization. Therefore, the training database should not only include responses to perturbations of the equilibrium structure such as strains and atomic displacements, but also snapshots of thermal vibrations at high temperatures. In order to avoid the potential fitting unnecessary phase space regions beyond thermal vibrations, we separately train the potential for each phase (HCP and BCC) of Zr studied in this work to ensure the accuracy of phonon dispersions. For both HCP and BCC Zr, the databases are constructed as follows.

**Database 1** is used to train the GAP model in the descriptor space around the equilibrium geometry and the mechanical response to bulk strains. Self-consistent field (SCF) calculations are performed with different strain tensors with distortion parameters up to 4% imposed on the simulation cell. The symmetry-irreducible strain tensors for the hexagonal lattice and the cubic lattice are specified in ref. [34] and ref. [35], respectively. The size of the simulation cells for HCP-Zr and BCC-Zr are specified in Table II. Database 1 includes forces on atoms, total energies and virial stress on the simulation cell.

**Database 2** is used to teach the GAP model with harmonic and anharmonic force constants at different volumetric strains. First, simulation cells of HCP-Zr and BCC-Zr are constructed with uniform strains on each lattice constant from -4% to 4% with the step of 1%. In each supercell with strains, small displacements (0.03 Å) are imposed to the irreducible atoms according to the space groups using the Phonopy package[36] and ShengBTE package[37]. SCF calculations are then performed for each perturbed supercell with strains and displacements, so that the total energies, forces, and virial stresses at the perturbed states are recorded.

**Database 3** provides the information of chemical environments and PES above 0 K. *ab-initio* molecular dynamics (AIMD) simulations were performed at different temperatures to generate snapshots of atomic configurations for both BCC and HCP Zr. At each temperature, 1000 snapshots of atomic configurations are generated with AIMD using a time step of 1 femtosecond. Total energy and forces are used as training data quantities.

All the training data in the databases is generated by the density functional theory (DFT) based first-principles calculations using the Vienna *Ab-initio* Simulation Package (VASP).[38, 39] Since the goal is to capture the effect of temperature on phonon dispersion (renormalization), the training database should include DFT data at both 0 K and at finite temperatures. All DFT calculations are performed using PBE functional[40] with projector augmented wave (PAW) method.[38, 39] For all DFT simulations, the cutoff energy is chosen as 300 eV.[6] For both HCP phase and BCC phase of Zr, the following databases were generated to train the GAP model with chemical environments. Table II summarizes the detailed parameters in DFT and AIMD calculations, including total number of atoms in all AIMD snapshots and DFT calculations ($N$), number of representative set of atoms ($M$), temperature $T$, dimensions of supercells, convergence threshold of SCF calculations (EDIFF tag in VASP package).

Table II. Detailed parameters for DFT calculations to generate training databases.

| | | | HCP Zr | | | |
|---|---|---|---|---|---|---|
| | $N$ | $M$ | T (K) | Supercell | k-mesh | EDIFF |
| Database 1 | 1350 | 20 | 0 | 3×3×2 | 7×7×7 | 1e-10 |
| Database 2 | 4266 | 65 | 0 | 3×3×2 | 7×7×7 | 1e-10 |
| Database 3 | 72000 | 750 | 100, 300 | 3×3×2 | 3×3×3 | 1e-6 |
| | | | BCC Zr | | | |
| | $N$ | $M$ | T (K) | Supercell | k-mesh | EDIFF |
| Database 1 | 1458 | 20 | 0 | 3×3×3 | 7×7×7 | 1e-10 |
| Database 2 | 2214 | 45 | 0 | 3×3×3 | 7×7×7 | 1e-10 |
| Database 3 | 54000 | 750 | 100, 300, 1200 | 3×3×3 | 3×3×3 | 1e-6 |

# 3. RESULTS AND DISCUSSIONS

This section discusses the application of GAP to model the phonon renormalization in Zr at elevated temperature. Before that, the accuracy of the GAP model to reproduce DFT calculations should be examined. As shown in Figure 1, the GAP prediction of total energies and components of forces ($F_{ix}, F_{iy}, F_{iz}$ of atom $i$ along three Cartesian axes) are compared with the original AIMD simulation, which are corresponding to 200 equally spaced snapshots randomly selected from the 1000 AIMD snapshots at 300 K. The GAP model is observed to reproduce the energies from AIMD calculation with the root mean squared error (RMSE) of 0.0002 eV/atom for the HCP phase and 0.0003 eV/atom for the BCC phase. The RMSE of the atomic forces between GAP model and AIMD simulations is 0.025 eV/Å for the HCP phase and 0.053 eV/Å for the BCC phase. The comparisons indicate good fitting of the *ab-initio* PES and its derivatives.

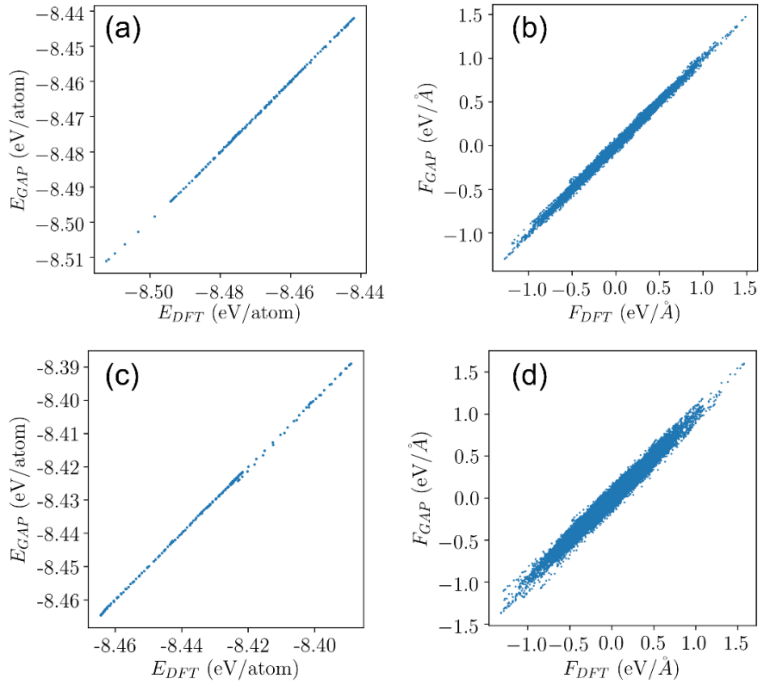

*Figure 1. (a-b). Comparison of (a) energy and (b) inter-atomic forces between GAP and AIMD calculations of the HCP-Zr. (c-d). Comparison of (c) energy and (d) force components between GAP and AIMD calculations of the BCC-Zr.*

In addition to accurately reproduce the training observables (energies and forces), the GAP model is also expected to reproduce the thermal and mechanical properties of the Zr crystals. Figure 2a shows the equation of state $E = E(V)$ and Figure 2b shows the symmetry-irreducible elastic constants $C_{ij}$ for both hcp and BCC-Zr. Excellent agreement is achieved in the equation of state as well as the elastic constants. The instability of the BCC-Zr is manifested in the elastic constants. For a crystal to be energetically stable, the Born criteria requires the $C_{ij}$ tensor to be positive-definite. In the case of BCC structure, the stability criteria requires $C_{11}, C_{12}$ and $C_{44}$ to satisfy $C_{11} - C_{12} > 0$, $C_{44} > 0$ and $C_{11} + 2C_{12} > 0$.[41] Clearly the criteria $C_{11} - C_{12} > 0$ is not satisfied as shown in the right panel of Figure 2b. Besides the elastic constants, we also compare the phonon dispersions predicted by the GAP model of both HCP-Zr and BCC-Zr at the static limit with the inelastic neutron scattering (INS) measurements[3, 42] and the DFT calculations, as shown in Figure 2c and Figure 2d. For the HCP phase, there is only small difference in the phonon dispersion, while larger discrepancy is observed for the soft modes (plotted as imaginary frequencies) of the BCC phase, which is likely due to the larger RMSE of energy and forces in for the BCC phase as shown in Figure 1b and 1d.

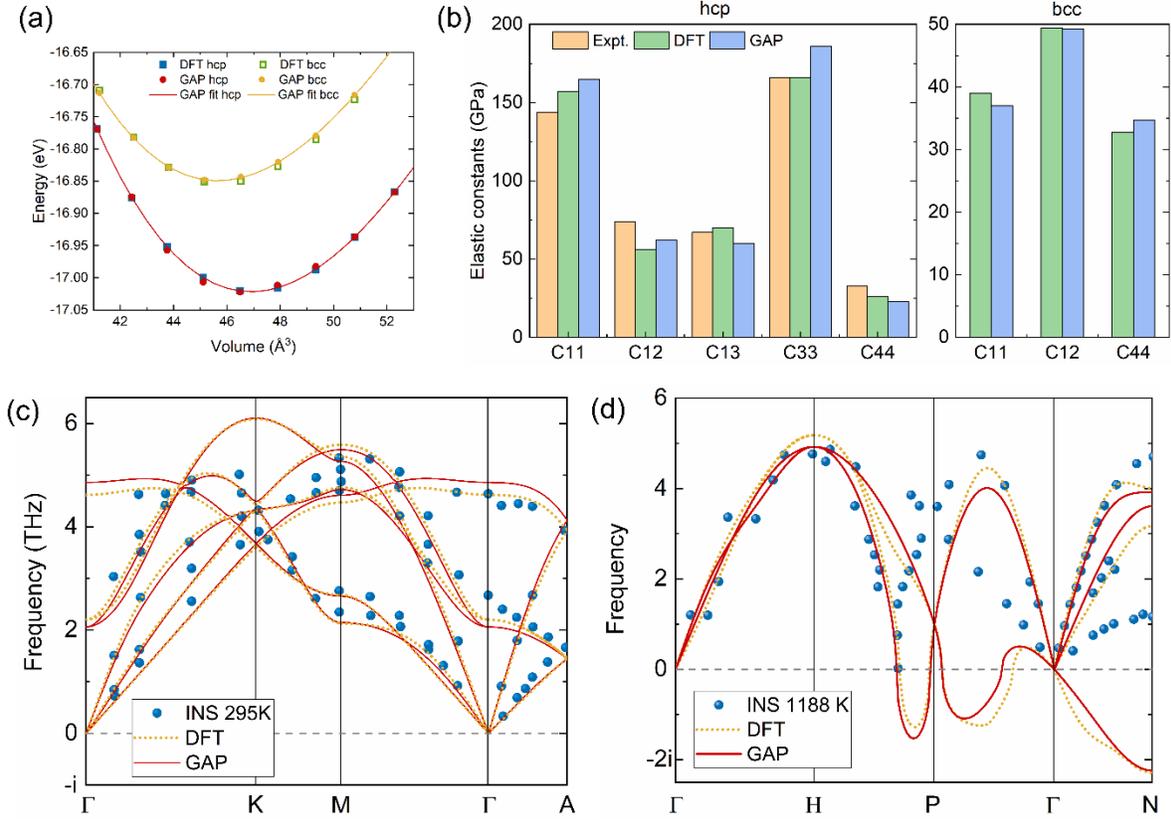

*Figure 2. (a) Equation of state (energy v.s. volume) of HCP-Zr and BCC-Zr calculated by GAP and DFT. (b) Symmetry-irreducible elastic constants of HCP-Zr (left panel) and BCC-Zr (right panel). The experimental elastic constants of HCP-Zr is from ref. [43] (c) Phonon dispersion of HCP-Zr. INS measurement data is taken from ref. [42] (d) Phonon dispersion of BCC-Zr. INS measurement data is taken from ref. [3]*

To illustrate the origin of the soft phonon modes, the PES is plotted in the normal coordinates for the two lowest modes at the high symmetry point N in the Brillouin zone. In order to obtain the shape of the PES around the equilibrium position, small displacements are imposed along the eigenvectors for the lowest soft TA mode with a scaling factor $Q_1$ and the second lowest TA mode with a scaling factor $Q_2$ and the PES as a function of scaled coordinates $E = E(Q_1, Q_2)$ is plotted as Figure 3a. It is clear that the PES shows a double-well shape. The dynamic instability of the BCC structure originates from the fact that the equilibrium state $(Q_1, Q_2) = (0,0)$ is a saddle point of the PES. Along the $Q_1$ direction, the equilibrium state is the local maxima of the double-well as shown in Figure 3b, while it is the local minima along the $Q_2$ direction. As a result of the negative local curvature $\frac{\partial^2 E}{\partial Q_1^2} < 0$, the eigenvalue for the lowest TA mode $\omega^2$ is also negative when the lattice dynamics simulations are performed at the static limit, so that the imaginary phonon frequency is observed in Figure 2d. At high temperature, the normal mode oscillator is hopping between the two potential wells, and the equilibrium position corresponding to the BCC structure is indeed the dynamical average between the two local minima. In addition, due to the complicated multi-minimum shape of the PES of the BCC phase, the fluctuations of AIMD energy and forces could also be larger compared with the stable HCP phase even at the same temperature, which results in the larger RMSE when reproducing the AIMD energies and forces.

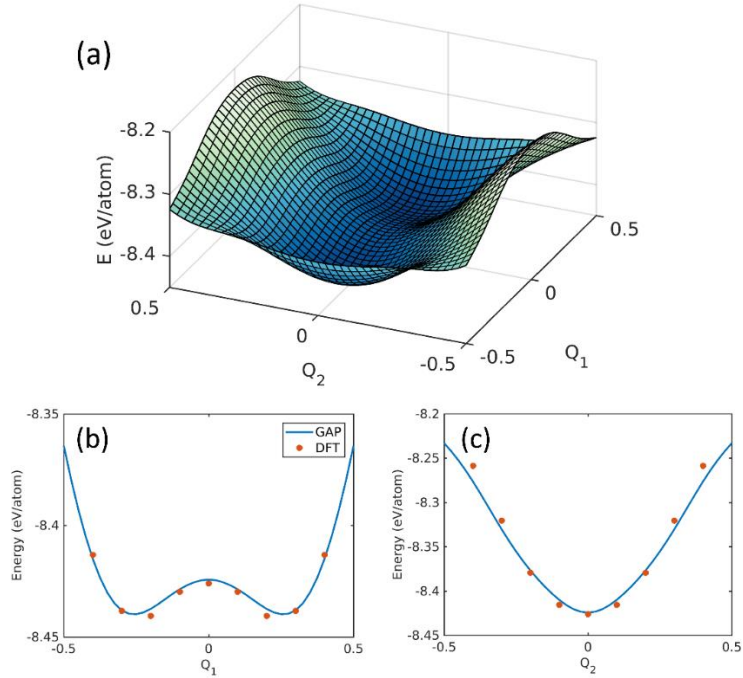

Figure 3. (a) PES along eigenvectors at high symmetry point N. $Q_1$ and $Q_2$ correspond to dimensionless normal coordinate of the two TA modes with the order of increasing frequency. (b) PES along the $Q_1$ direction with $Q_2 = 0$. (c) PES along the $Q_2$ direction with $Q_1 = 0$.

With the idea that the BCC structure is stabilized through dynamical average of the low-symmetry minima of the PES, the phonon dispersion should be renormalized to real frequency values at high temperature when the PES is dynamically sampled. MD simulations are therefore performed to stochastically sample the PES, using the machine learning driven GAP potentialas we have developed above. Phonon dispersion is then calculated by SED analysis[44, 45] which maps the vibrational energy distribution in wave-vector space and frequency domain $(\boldsymbol{q}, \omega)$. Here the SED distribution is calculated by summing the Fourier transform of the amplitudes of vibrational velocities:

$$\phi(\boldsymbol{q},\omega) = \frac{1}{4\pi\tau_0 N_{cells}} \sum_{\alpha=x,y,z} \sum_{b} m_b \left| \int_0^{\tau_0} \sum_{\boldsymbol{R}} \dot{u}_\alpha(\boldsymbol{R},b,t) \cdot \exp(i\boldsymbol{q}\cdot\boldsymbol{R} - i\omega t)\, dt \right|^2 \quad (10)$$

where $N_{cells}$ is the total number of unit cells, $\boldsymbol{R}$ is the lattice vector and $b$ is the index of basis atoms in the unit cell, $\dot{u}_\alpha(\boldsymbol{R},b,t)$ is the velocity component along the $\alpha = (x,y,z)$ axis of the atom ($\boldsymbol{R}$,b) at time $t$. The quantity $dt$ (=1 fs) is the time step between neighboring MD snapshots, and $\tau_0$, the total integration time is selected as 1 ns, and longer $\tau_0$ is found not to affect the SED distributions. For the HCP phase, SED along the $\Gamma - A$ direction and the $\Gamma - M$ direction are calculated, using supercells containing 3×3×50 primitive cells and 50×3×3 primitive cells, respectively. For the BCC phase, SED is extracted along the $\Gamma - N$ path using a supercell containing 50×3×3 primitive cells. Figure 4a-b shows the SED of the HCP-Zr at 100 K and 300 K. For the HCP phase, the most pronounced effect of non-zero temperature is the broadening of the SED lines due to stronger phonon scattering at higher temperature. Figure 4c shows the phonon dispersion of the BCC-Zr at 1188 K. The SED analysis has successfully captured the renormalization of the soft TA mode in BCC-Zr which is now renormalized to ~1 THz at 1188 K.

Figure 4d shows the SED as a function of frequency at $q$=(0.3, 0, 0) along the Γ − N direction. The broad SED peak observed near 1 THz is agreeing well with INS experiments.[3]

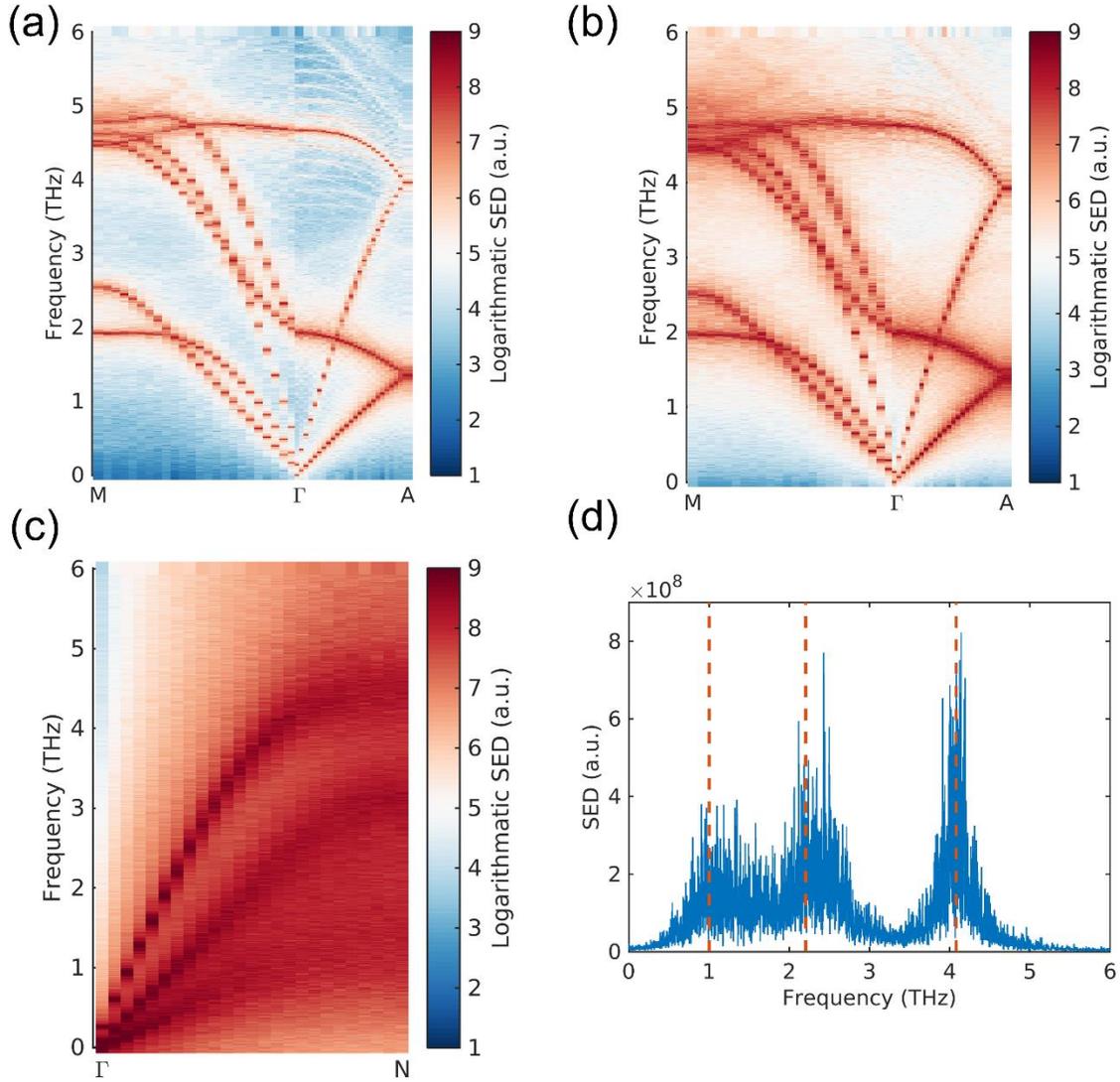

Figure 4. (a-b) SED of HCP-Zr at (a) 100 K and (b) 300 K. (c) SED of bcc-Zr at 1188 K. (d) SED as a function of phonon frequency at $q$ = (0.3,0,0). The dashed lines indicate the frequency measured by INS in ref.[3] at 1188 K.

# 4. SUMMARY

In summary, we studied the temperature effect on phonon dispersions of the HCP phase and the dynamically unstable BCC phase of Zr, using molecular dynamics simulation with machine learning-driven Gaussian approximation potential. The GAP model accurately reproduces energies and interatomic forces corresponding to the atomic configurations of the AIMD snapshots as well as the mechanical properties of Zr. The dynamical instability of BCC Zr is captured by the GAP model with the soft phonon modes in the dispersion relationship as well as the non-positive-definite elastic constant tensor. The instability of the BCC structure is observed to originate from the double-well shape of the PES, and the BCC phase becomes stable at high temperature as a result of dynamical average as the normal mode oscillators hopping between the two local minima of the PES. The stabilization of BCC Zr is captured by examining the phonon dispersion at high temperature using MD simulations and SED analysis. In addition to the broadening effect at elevated temperature, the SED analysis also captures the phonon renormalization of the soft TA mode in BCC crystal, with the frequency renormalized to ~ 1THz at 1188 K, agreeing well with the INS experiments. This work for the first time approaches the problem of phonon renormalization in dynamically unstable crystals using molecular dynamics, showing that machine learning-driven potential is a promising tool for modeling high temperature lattice dynamics and thermal properties.

**Acknowledgement:** This work is supported by NSF (Grant No. 1512776). All calculations are performed using the Summit supercomputer, which is supported by NSF (awards ACI-1532235 and ACI-1532236), University of Colorado Boulder and Colorado State University. X.Q. acknowledges helpful discussions with Xinpeng Zhao.

**References:**

1. G. Kresse, J. Furthmuller and J. Hafner, Europhys. Lett. **32** (9), 729-734 (1995).
2. Stefano Baroni, S. d. Gironcoli and A. D. Corso, Reviews of Mordern Physics **73**, 515 (2001).
3. A. Heiming, W. Petry, J. Trampenau, M. Alba, C. Herzig, H. R. Schober and G. Vogl, Phys. Rev. B **43** (13), 10948-10962 (1991).
4. W. Petry, A. Heiming, J. Trampenau, M. Alba, C. Herzig, H. R. Schober and G. Vogl, Phys. Rev. B **43** (13), 10933-10947 (1991).
5. D. J. Hooton, Z. Phys. **142**, 42-58 (1955).
6. P. Souvatzis, O. Eriksson, M. I. Katsnelson and S. P. Rudin, Phys. Rev. Lett. **100** (9), 095901 (2008).
7. O. Hellman and I. A. Abrikosov, Phys. Rev. B **88** (14), 144301 (2013).
8. S.-Y. Yue, X. Zhang, G. Qin, S. R. Phillpot and M. Hu, Phys. Rev. B **95** (19), 195203 (2017).
9. T. Feng, L. Lindsay and X. Ruan, Phys. Rev. B **96** (16), 161201(R) (2017).
10. X. Qian, X. Gu and R. Yang, Nano Energy **41**, 394-407 (2017).
11. A. Rohskopf, H. R. Seyf, K. Gordiz, T. Tadano and A. Henry, npj Computational Materials **3** (1) (2017).
12. P. C. Howell, The Journal of chemical physics **137** (22), 224111 (2012).
13. J. Behler, The Journal of chemical physics **145** (17), 170901 (2016).
14. C. M. Bishop, *Neural Networks for Pattern Recognition*. (Oxford University Press, Oxford, 1995).
15. C. E. Rasmussen and C. K. I. Williams, *Gaussian Processes for Machine Learning*. (MIT Press, 2006).
16. A. Seko, A. Takahashi and I. Tanaka, Phys. Rev. B **92** (5), 054113 (2015).
17. J. Behler and M. Parrinello, Phys. Rev. Lett. **98** (14), 146401 (2007).
18. A. P. Bartok, M. C. Payne, R. Kondor and G. Csanyi, Phys. Rev. Lett. **104** (13), 136403 (2010).
19. A. P. Bartók, J. Kermode, N. Bernstein and G. Csányi, ArXiv:1085.01568 (2018).
20. P. Rowe, G. Csányi, D. Alfè and A. Michaelides, Phys. Rev. B **97** (5), 054303 (2018).
21. V. L. Deringer and G. Csányi, Phys. Rev. B **95** (9), 094203 (2017).
22. K. Miwa and H. Ohno, Physical Review Materials **1** (5), 053801 (2017).
23. S. Fujikake, V. L. Deringer, T. H. Lee, M. Krynski, S. R. Elliott and G. Csanyi, The Journal of chemical physics **148** (24), 241714 (2018).
24. G. C. Sosso, G. Miceli, S. Caravati, J. Behler and M. Bernasconi, Phys. Rev. B **85** (17), 174103 (2012).
25. Kharchenko and Kharchenko, Condens. Matter Phys. **16** (1), 13801 (2013).
26. H. Zong, G. Pilania, X. Ding, G. J. Ackland and T. Lookman, Npj Computational Materials **4**, 48 (2018).
27. A. Bartók-Pártay, *The Gaussian Approximation Potential: an interatomic potential derived from first principles quantum mechanics*. (Springer Theses, 2010).
28. *See Supplemental Material at [URL will be inserted by publisher] for the training databases and developed GAP potential files for HCP and BCC Zr.*

29. *The QUIP package is open to academic users at http://www.libatoms.org/Home/Software*.
30. A. P. Bartók and G. Csányi, Int. J. Quantum Chem **115** (16), 1051-1057 (2015).
31. D. Dragoni, T. D. Daff, G. Csányi and N. Marzari, Physical Review Materials **2** (1), 013808 (2018).
32. A. P. Bartók, R. Kondor and G. Csányi, Phys. Rev. B **87** (18), 184115 (2013).
33. W. J. Szlachta, A. P. Bartók and G. Csányi, Phys. Rev. B **90** (10), 104108 (2014).
34. L. Fast, J. M. Wills, B. Johansson and O. Eriksson, Phys. Rev. B **51** (24), 17431-17438 (1995).
35. M. J. Mehl, Phys. Rev. B **47** (5), 2493-2500 (1993).
36. A. Togo and I. Tanaka, Scripta Mater. **108**, 1-5 (2015).
37. W. Li, J. Carrete, N. A. Katcho and N. Mingo, Comput. Phys. Commun. **185** (6), 1747-1758 (2014).
38. G. Kresse and J. Furthmuller, Comput. Mater. Sci. **6**, 15-50 (1996).
39. G. Kresse and D. Joubert, Phys. Rev. B **59** (3), 1758 (1991).
40. J. P. Perdew, K. Burke and M. Ernzerhof, Phys. Rev. Lett. **77**, 3865 (1996).
41. M. Born, Cambridge Philosophical Society **36**, 160 (1940).
42. C. Stassis, J. Zarestky, D. Arch, O. D. McMasters and B. N. Harmon, Phys. Rev. B **18** (6), 2632-2642 (1978).
43. E. A. Brandes, *Smithells Metals Reference Book, 6th ed.* . (Butterworth, London, 1983).
44. J. A. Thomas, J. E. Turney, R. M. Iutzi, C. H. Amon and A. J. H. McGaughey, Phys. Rev. B **81** (8), 081411(R) (2010).
45. T. Feng, B. Qiu and X. Ruan, J. Appl. Phys. **117** (19), 195102 (2015).